\PassOptionsToPackage{unicode}{hyperref}
\PassOptionsToPackage{hyphens}{url}

\documentclass[]{article}

\usepackage[top=1.3in, bottom=1.3in, left=1.3in, right=1.3in]{geometry}

\usepackage{amsmath,amssymb}
\usepackage{lmodern}
\usepackage{iftex}
\ifPDFTeX
  \usepackage[T1]{fontenc}
  \usepackage[utf8]{inputenc}
  \usepackage{textcomp} % provide euro and other symbols
\else % if luatex or xetex
  \usepackage{unicode-math}
  \defaultfontfeatures{Scale=MatchLowercase}
  \defaultfontfeatures[\rmfamily]{Ligatures=TeX,Scale=1}
\fi
\IfFileExists{upquote.sty}{\usepackage{upquote}}{}
\IfFileExists{microtype.sty}{% use microtype if available
  \usepackage[]{microtype}
  \UseMicrotypeSet[protrusion]{basicmath} % disable protrusion for tt fonts
}{}
\makeatletter
\@ifundefined{KOMAClassName}{% if non-KOMA class
  \IfFileExists{parskip.sty}{%
    \usepackage{parskip}
  }{% else
    \setlength{\parindent}{0pt}
    \setlength{\parskip}{6pt plus 2pt minus 1pt}}
}{% if KOMA class
  \KOMAoptions{parskip=half}}
\makeatother
\usepackage{xcolor}
\usepackage{graphicx}
\makeatletter
\def\maxwidth{\ifdim\Gin@nat@width>\linewidth\linewidth\else\Gin@nat@width\fi}
\def\maxheight{\ifdim\Gin@nat@height>\textheight\textheight\else\Gin@nat@height\fi}
\makeatother

\setkeys{Gin}{width=\maxwidth,height=\maxheight,keepaspectratio}
\makeatletter
\def\fps@figure{htbp}
\makeatother
\ifLuaTeX
  \usepackage{selnolig}  % disable illegal ligatures
\fi
\ifPDFTeX
  \newcommand{\RL}[1]{\beginR #1\endR}

\fi
\IfFileExists{bookmark.sty}{\usepackage{bookmark}}{\usepackage{hyperref}}
\IfFileExists{xurl.sty}{\usepackage{xurl}}{} 
\hypersetup{
  pdftitle={Influence of tip materials on the friction force microscopy of epitaxial graphene on SiC(0001): comparison of diamond and silicon tips in experiments and atomistic simulations},
  hidelinks,
  pdfcreator={LaTeX via pandoc}}

\title{\textbf{Influence of tip materials on the friction force
microscopy of epitaxial graphene on SiC(0001): comparison of diamond and
silicon tips in experiments and atomistic simulations}}
\author{}

\begin{document}
\maketitle

\begin{center}
    \large \textbf{Mohammad Zarshenas}\textsuperscript{1,2}, 
    \textbf{Takuya Kuwahara}\textsuperscript{3}, 
    \textbf{Bartosz Szczefanowicz}\textsuperscript{4}, 
    \textbf{Andreas Klemenz}\textsuperscript{1}, 
    \textbf{Lars Pastewka}\textsuperscript{5}, 
    \textbf{Gianpietro Moras}\textsuperscript{1}, 
    \textbf{Roland Bennewitz}\textsuperscript{6}, 
    \textbf{Michael Moseler}\textsuperscript{1,2*}
\end{center}

\vspace{1em}

\begin{center}
    \textit{\textsuperscript{1}Fraunhofer IWM, MikroTribologie Centrum µTC, 
    Wöhlerstraße 11, 79108 Freiburg, Germany} \\

    \textit{\textsuperscript{2}Institute of Physics, University of Freiburg, 
    Hermann-Herder-Straße 3, 79104 Freiburg, Germany} \\

    \textit{\textsuperscript{3}Department of Mechanical Engineering, Osaka 
    Metropolitan University, 3-3-138 Sugimoto, Osaka, 558--8585, Japan} \\

    \textit{\textsuperscript{4}Marian Smoluchowski Institute of Physics, 
    Jagiellonian University, 30-348 Krakow, Poland} \\

    \textit{\textsuperscript{5}Department of Microsystems Engineering, University of 
    Freiburg, Georges-Köhler-Allee 103, 79110 Freiburg, Germany} \\

    \textit{\textsuperscript{6}INM--Leibniz-Institute for New Materials and Physics 
    Department, Saarland University, 66123 Saarbrücken, Germany}
\end{center}

\vspace{1em}

\ifnum\value{page}=1
    \begin{center}
        \small *Contact author: michael.moseler@iwm.fraunhofer.de
    \end{center}
\fi

\vspace{1em}
\begin{abstract}
Friction force microscopy (FFM) with silicon tips on epitaxial graphene
supported by SiC(0001) previously revealed a sharp increase in friction
at a threshold normal force, linked to the intermittent rehybridization
of graphene and the formation of single-layer diamond above 12 GPa. In
this study, the FFM behavior of a diamond tip is compared with that of
the silicon tip. The diamond tip exhibited a similar abrupt increase in
friction, but the threshold normal force was approximately twice as high
for comparable tip radii. Simulations of graphene on SiC(0001) sliding
against hydroxylated amorphous carbon (a-C) and silicon oxide
(SiO\textsubscript{2}) slabs show that both systems exhibit low shear
stress at low pressures, which increases at higher pressures due to bond
formation between the graphene and counter slabs. The pressure
dependence differs slightly: SiO\textsubscript{2} reaches a shear stress
plateau around 6 GPa, while a-C continues to increase gradually. For
a-C, the transition threshold shifts to higher pressures, consistent
with FFM results. Temperature lowers the transition threshold
significantly, while sliding velocity has minimal impact on shear
stress. These findings provide insights into the stability of
low-friction interfaces between epitaxial graphene and the key materials
which come into contact with graphene in current
micro-electro-mechanical systems.
\end{abstract}

\section{INTRODUCTION}

Flat graphene layers are known to exhibit very low friction. The strong
covalent in-plane C-C bonds provide graphene with high stability against
chemical and mechanical wear. Therefore, graphene is an intriguing 2D
model system to investigate mechanisms which can lead to ultra-low
friction ~{[}1--4{]}. Epitaxial graphene on SiC(0001) exhibits super-low
friction due to its weak out-of-plane interactions ~{[}5{]}. Using FFM,
Szczefanowicz et al. studied the tribological behavior of epitaxial
graphene using silicon tips with SiO\textsubscript{2} surface layer,
revealing a sudden transition between a low and high friction regime
~{[}6{]}. They found with accompanying DFTB simulations that the
threshold pressure for entering the high-friction regime was 12.7 GPa
and that the atomic configuration of the amorphous SiO\textsubscript{2}
played a crucial role in determining the covalent bond formation in the
transition regime. The study highlighted the importance of understanding
also chemical aspects in the tribological behavior of graphene across
various sliding regimes.

While this previous research focused on the sliding behavior between a
silicon tip and epitaxial graphene ~{[}6{]}, the present study aims to
investigate the sliding interaction between epitaxial graphene on
SiC(0001) and a diamond tip covered by an amorphous carbon surface
layer. The use of a diamond tip provides an alternative material system
for exploring the mechanisms leading to ultra-low friction, as the
bonding mechanisms and material properties of the a-C differ from those
of the SiO\textsubscript{2} surface layer. By comparing the behavior of
a diamond and a silicon tip, we aim to elucidate the influence of
different bonding mechanisms and material properties on sliding behavior
and shear stresses. Overall, this investigation contributes to a deeper
understanding of the mechanisms governing the sliding interaction
between a-C covered diamond and epitaxial graphene, shedding light on
the factors influencing friction and providing valuable insights for the
design of low-friction interfaces -- for instance in modern
micro-electro-mechanical systems (MEMS) applications.

\section{METHODS}

\subsection{Experiment}

The friction between epitaxial graphene on SiC(0001) and a diamond tip
was measured using friction force microscopy (FFM). The lateral force
acting on the tip as it slides over the graphene at different normal
forces was recorded. The experimental procedures for the atomic force
microscopy (AFM) experiments were described in Ref. ~{[}6{]}. Briefly,
the graphene/SiC(0001) samples were grown by thermal decomposition
~{[}7{]} and purified by annealing in ultra-high vacuum (UHV) prior to
the experiments. Lateral forces were recorded in UHV by scanning the FFM
tip in contact with atomically flat regions of monolayer graphene and
measuring the torsion of the FFM cantilever caused by the lateral force.
Friction was quantified as the mean value of lateral forces for scan
areas that were large compared to the SiC(0001) surface structure. A
diamond tip (Nanosensors - Adama Innovations) was used with an estimated
initial radius of 5 nm. The results of this FFM study will be compared
to the already published FFM experiments that used a silicon tip with a
similar radius ~{[}6{]}.

\subsection{Simulations}

Molecular dynamics (MD) simulations were conducted using the
self-consistent-charge density functional tight-binding (DFTB) method
~{[}8{]} within the framework of the ATOMISTICA suite ~{[}9{]}. The
substrate consisted of graphene and 6 layers of SiC(0001) structure with
dimensions 1.07×0.92×6.0 nm³. An a-C counter surface was designed to
conform to these dimensions, with a density of 2.2 g/cm³ and 163 carbon
atoms. The a-C structure was obtained by solidifying molten a-C under
periodic boundary conditions (PBC) applied along all three Cartesian
axes, with the system undergoing quenches at constant volume. Initially,
the a-C was quenched to 2000 K for 25 ps, followed by another step of
quenching to 300 K for an additional 25 ps, with stress relaxation steps
incorporated during each stage while maintaining a fixed cell size.
Subsequently, the PBC along the z-direction were removed, creating two
free surfaces (an upper and a lower one). The undercoordinated carbon
atoms on the upper surface of the a-C were passivated with hydrogen
atoms, while the lower surface underwent chemisorption through the
introduction of eight water molecules during pressure equilibration.
This process involved the adsorption of water molecules, leading to the
incorporation of hydrogen atoms and hydroxyl groups. The top atoms of
the a-C slab within a thickness of 0.5 nm, as well as a 0.5 nm bottom
layer of the graphene/SiC slab were maintained rigid. Detailed
simulation parameters for a comparable system with a
SiO\textsubscript{2} counter surface are provided in a previous
publication (Ref. ~{[}6{]}).

During the sliding MD simulations, the Pastewka-Moser-Moseler
pressure-coupling algorithm ~{[}10{]} was employed. The rigid a-C atoms
underwent sliding motion along the \(x\)-axis at velocities of 10 m/s
and 100 m/s, operating under varying normal pressures ranging from 5 GPa
to 22.5 GPa, while the positions of the rigid layers of the graphene/SiC
remained unchanged. A Langevin thermostat ~{[}11{]}, acting
perpendicular to the sliding direction, regulated the constant
system\textquotesingle s temperatures of 300 K, 500 K and 1000 K. The
equations of motion were integrated using the velocity Verlet algorithm
~{[}11{]} with a time step of 0.5 fs. To calculate the shear stress, the
force components exerted along the \(x\)-axis on the rigid layers of the
a-C were summed and then divided by the lateral area of the simulation
cell. Subsequently, the calculated shear stress was averaged over the
final half of the total simulation time of 0.2 ns.

To analyze the bonding behavior in the system, we calculated the number
of bonds between specific atom pairs across multiple configurations over
the course of our 0.2 ns simulations. Bonds were identified based on
interatomic distances, using a threshold of 1.85 Å for C-C bonds and 2.2
Å for C-Si bonds. This process was applied to every configuration in the
simulation, systematically evaluating all atomic pairs to determine bond
counts in each frame. To ensure statistical robustness, the analysis was
performed on four independent samples, each represented by an
independent simulation trajectory.

The mass distribution was examined by calculating the density profile
along the \(z\)-axis of the simulation box. The simulation box was
divided into discrete 0.4 Å bins along the \(z\)-direction, and atoms
were assigned to these bins based on their \(z\)-coordinates. For each
bin, the total mass of atoms was computed by summing the atomic masses
of all atoms within the bin using predefined values for each atomic
species. To calculate the velocity profile, the simulation box was
similarly divided into layers along the \(z\)-direction, grouping
particles within each layer. The \(x\)-velocities of these particles in
the sliding direction were then averaged to provide a representative
velocity for each layer. The calculated velocity profile was averaged
over the final half of the total simulation time.

\section{RESULTS}

\subsection{Experiment}

We compare our FFM experiments using a diamond tip with previously
published results {[}1{]} from an identical study using a silicon tip.
Fig. 1 displays the friction force acting on a Si tip (initially and
after repeated FFM scans -- see green disks and black squares) as a
function of normal pressure and compares it with the diamond tip FFM
experiment (red squares). For the Si tip, Szczefanowicz and coworkers
~{[}6{]} observed an initial linear low-friction regime up to a force of
80~nN, followed by an abrupt increase in the average slope of the
friction versus normal pressure curve. High-friction values generally
increase with pressure but exhibit significant scatter, occasionally
returning to the low-friction regime. After repeating the friction force
experiment three more times the transition to high friction occurs at a
higher normal force of 190~nN (4th series in Fig. 1). This effect was
explained by blunting of the Si tip. Atomistic simulations were
performed to explore the shear response between epitaxial graphene and
the SiO\textsubscript{2} surface layer of the Si tip. They revealed that
the step-like increase in shear stress is directly related to the
formation of chemical bonds between the graphene layer and the oxygen or
silicon atoms in the silicon oxide as well as the intermittent formation
of a single diamond layer ~{[}6{]}.

\begin{figure}[htbp]
\centering
\includegraphics[width=0.8\textwidth,height=0.5\textheight,keepaspectratio]{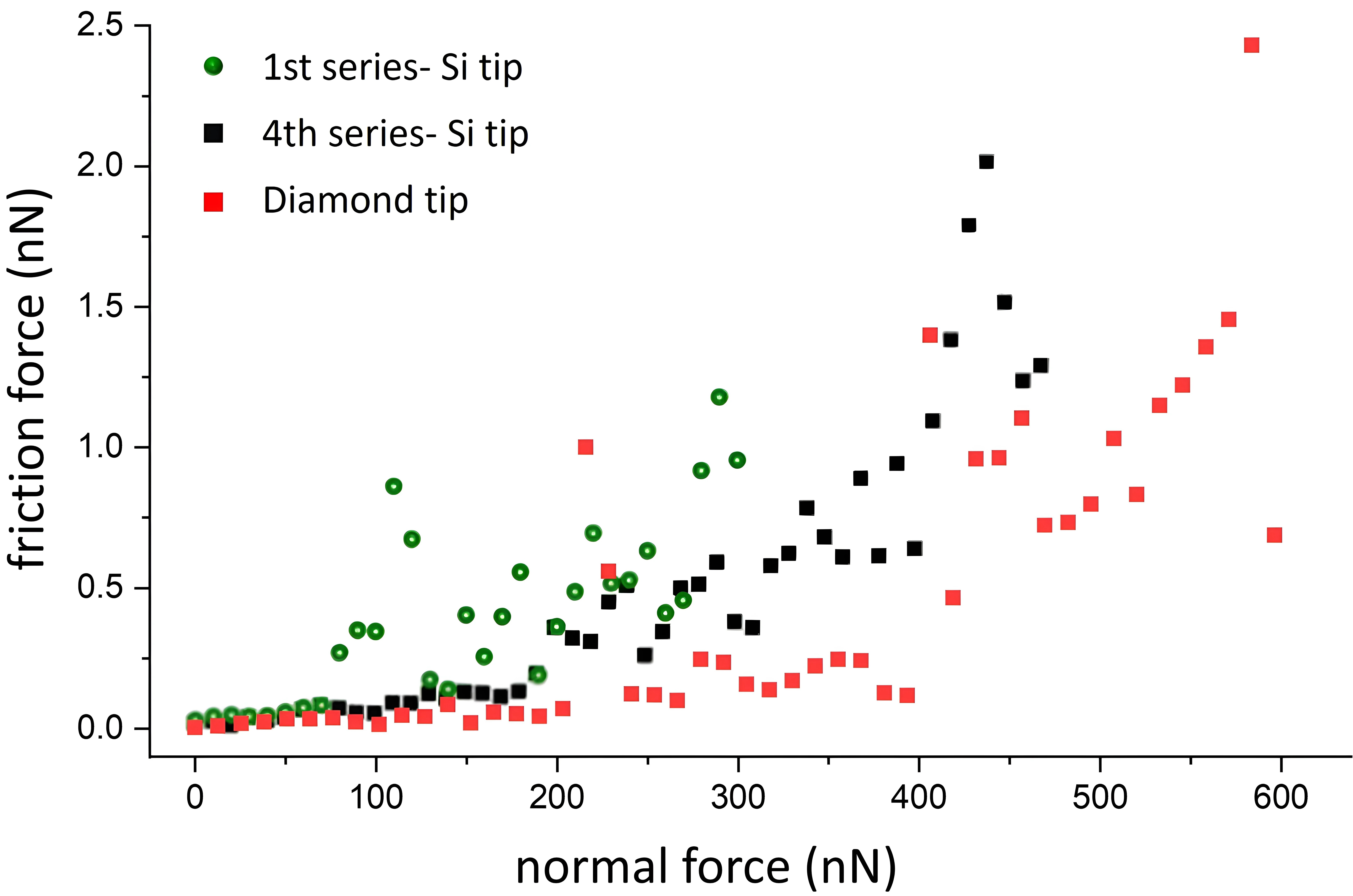}
\caption{Friction force as a function of the applied normal force recorded
with a silicon and a diamond tip sliding over epitaxial graphene on a
SiC(0001) substrate.}
\label{fig:1}
\end{figure}

Low friction with a linear increase as a function of applied normal
force and with fluctuations in the magnitude of the frictional force was
observed for the diamond tip for normal forces up to 400 nN. Beyond 400
nN there is a sudden increase in the friction force, reaching up to 2.5
nN. Please note that the graphene did not rupture during the experiments
with the diamond tip. The integrity of the graphene layer was confirmed
by the observation of periodic atomic-scale stick-slip pattern in the
measured lateral force. Experiments performed at loads higher than those
here show that rupture of the graphene leads to large irregular
fluctuations in the lateral force signal (see supplementary material).
Due to the sudden increase in the slope of the friction force curve with
respect to the normal force, it can be hypothesized that the pressure
exerted by the diamond tip also causes the formation of a single diamond
layer (i.e., sp\textsuperscript{2}-to-sp\textsuperscript{3}
rehybridization of graphene, analogous to the transformation caused by
the silicon tip) and the bonding of this layer to an a-C layer covering
the diamond tip. In the following section, we confirm this hypothesis
using DFTB simulations.

\subsection{Simulations}

\subsubsection{Sliding regimes}

Molecular dynamics simulations were conducted to model the sliding
interaction between a passivated a-C slab (representing the a-C
overlayer on the diamond tip) and an epitaxial graphene sheet on
SiC(0001). To account for local variations in the amorphous carbon
overlayer on the diamond tip, four different structural a-C
configurations were prepared to simulate diverse atomic arrangements,
following the same approach used for the SiO\textsubscript{2} slabs in
Ref. ~{[}6{]}. The simulations aim to investigate the pressure
dependence of the shear behavior and the possible subsequent structural
changes at the interface between the diamond tip and the epitaxial
graphene. Fig. 2(a) displays representative snapshots captured during
the final 0.1 ns of a 0.2 ns simulation period. The velocity profiles
(green curves in Fig. 2(a)) are superimposed in order to mark the
spatial region that accommodates the applied shear. The a-C slab
simulations are compared with the SiO\textsubscript{2} slab results from
Ref. {[}1{]} -- see the lowest row in Fig. 2(a). By applying a
predefined normal pressure between 5 and 22.5 GPa, the simulations
systematically evaluate the structural evolution at the interface
between the amorphous layers and graphene/SiC(0001) during sliding at a
temperature of 300 K and a velocity of 100 m/s. While Fig. 2(a) reports
representative results from one of the four trajectories, Fig. 2(b)
displays the mean steady state shear stress of all four runs as a
function of applied normal pressure (black symbols for a-C and red
symbols for SiO\textsubscript{2}). An inspection of Fig. 2(a) and (b)
reveals that the tribo systems with the a-C and the SiO\textsubscript{2}
slabs exhibited distinct sliding regimes, with different characteristics
emerging across the examined normal pressure range.

\begin{figure}[htbp]
\centering
\includegraphics[width=\textwidth,height=0.8\textheight]{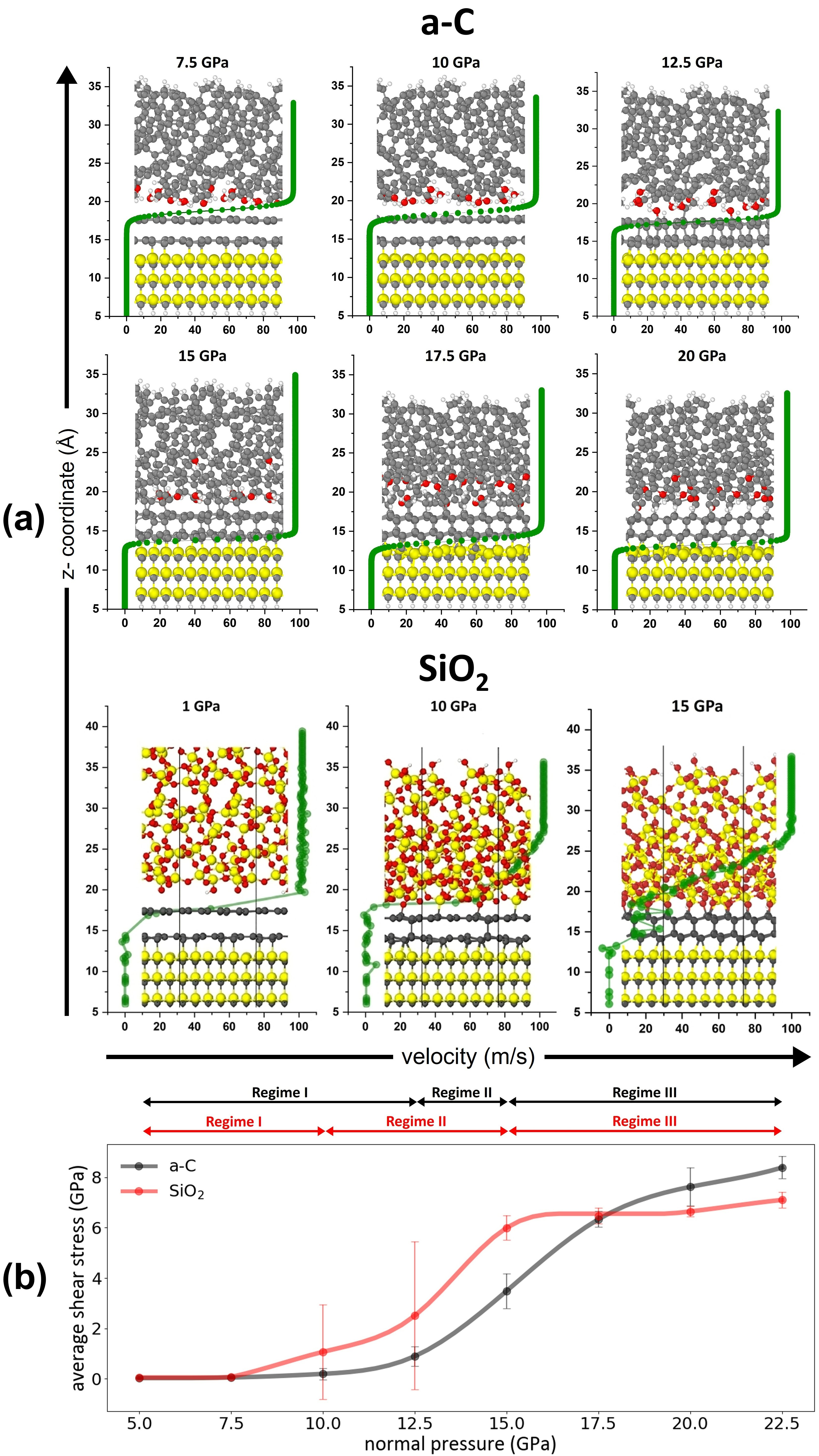}
\caption{Atomistic simulation of an a-C and SiO\textsubscript{2} slab
sliding on epitaxial graphene/SiC(0001) with a velocity of 100 m/s and
at a temperature of 300 K. (a) Representative snapshots that were
captured during the final 0.1 ns of a 0.2 ns simulation period. Colors
distinguish elements: yellow for silicon, red for oxygen, gray for
carbon, and white for hydrogen. Sticks between spheres denote chemical
bonds. The green data points show the sliding velocity as a function of
the normal coordinate (\(z\)) and mark the location of the shear plane.
(b) Average shear stress versus normal pressure for a-C and
SiO\textsubscript{2} (each averaged over 4 different runs). \RL{}The
arrows at the top of the figure indicate the approximate range of the
normal pressures at which each sliding regime occurs, with black arrows
for a-C and red arrows for SiO\textsubscript{2}. Solid lines represent
spline fits to the data points to guide the eye.}
\label{fig:2}
\end{figure}

\emph{Regime I}. For normal pressure less than 12.5 GPa, no chemical
bonds are formed between the a-C and graphene, nor between graphene and
the carbon interface layer (IFL) on top of the SiC. In this regime, the
shear plane is located between the H/OH terminated a-C and graphene.
Similarly, for the SiO\textsubscript{2} slab, no chemical bonds form
between the interfaces for normal pressures below 10 GPa, and the shear
plane is located between the SiO\textsubscript{2} and graphene.
Consequently, the shear stress in both cases remains small (for SiO\textsubscript{2}: \(\tau < 0.061~\text{GPa}\) and for a-C: \(\tau < 0.045~\text{GPa}\) (see Fig.~2(b)) and the resulting friction coefficients (for SiO\textsubscript{2}: \(\mu = \frac{\tau}{P} = 0.008 < 0.01\), for a-C: \(\mu = \frac{\tau}{P} = 0.006 < 0.01\)) indicate superlubricious sliding.

\emph{Regime II}. As the normal pressure reaches 12.5 GPa, C-C bonds
form between the a-C and graphene, as well as between graphene and the
IFL, marking the onset of a new sliding regime. This observation aligns
with Bundy et et al.'s findings that graphite transitions to diamond at
pressures up to approximately 13 GPa ~{[}12{]}. In this regime, the
shear plane remains in the gap between the a-C and graphene, gradually
shifting closer to the graphene layer. In addition, the terminal H and
OH groups transfer from the a-C tip to the graphene and some ether
functional groups form on the graphene that bond back with the a-C at
this pressure range. The SiO\textsubscript{2} tip enters a similar
regime at a normal pressure of 10 GPa, where
SiO\textsubscript{2}-graphene and graphene-IFL bond formation begins. In
these cases, however, the shear plane moves upwards due to plastic
events in the SiO\textsubscript{2} in addition to the sliding at the
graphene-SiO\textsubscript{2} interface.

\emph{Regime III}. As the normal pressure increases to 15 GPa, C-C bond
formation increases between the terminated a-C and graphene, as well as
between graphene and the IFL. The increased bonding between graphene and
IFL causes them to be dragged along with the a-C, resulting in a new
sliding regime where the shear plane is predominantly located between
the IFL and SiC. This regime persists up to our maximum normal pressure
of 22.5 GPa. Notably, at this pressure range, the terminating H and OH
groups mostly migrate away from the graphene and bond with carbon atoms
deeper within the a-C, playing a minor role at the a-C/graphene
interface. Similarly, at a normal pressure of 15 GPa for the
SiO\textsubscript{2} slab, behavior comparable to that observed with a-C
was noted, with an almost full connection between graphene and the IFL.
Moreover, numerous bonds (O-C and Si-C) formed between
SiO\textsubscript{2} and graphene, resulting in the emergence of sliding
regime III, where the shear plane now shifts fully upward into the
silicon oxide. In addition, this regime continued for normal pressures
up to 22.5 GPa. A comparison of measured normal and friction forces with
calculated normal pressure and shear stress requires a contact mechanics
model, which critically depends on configurational details at the tip
apex and on the velocity dependence of the friction forces. We refer
readers to Ref. {[}1{]} for a quantitative estimate and further
discussion.

The different shear plane locations observed in the a-C and
SiO\textsubscript{2} simulations can be attributed to the different
mechanical properties of both materials. The strong covalent C-C bonds
in a-C contribute to its high hardness and resistance to deformation,
allowing the material to maintain its rigidity. This rigidity, combined
with robust bonding between a-C and graphene, ensures that the
graphene/IFL group remains firmly attached to the a-C, thereby
positioning the shear plane at the interface between the IFL and SiC. In
contrast, the SiO\textsubscript{2} structure, which involves Si-O bonds,
is relatively soft and can form strong Si-C and O-C bonds with the
graphene layer (which has evolved to a single layer of diamond). This
softer nature along with the strong bond formation at the interface
causes the shear plane to shift upward into the silicon oxide layer
under increasing normal pressures. The impact of bond formation on shear
stress across different interfaces will be explored in detail in the
next section.

In the following, we discuss the variation of shear stress for the a-C
and SiO\textsubscript{2} systems under different normal pressures (Fig.
2(b)). Our results show that the shear stress for the a-C significantly
increases with pressure. At lower contact pressures, the a-C
demonstrates smooth sliding over graphene with minor shear stress due to
the absence of chemical bonding (see 7.5 and 10 GPa snapshots in Fig.
2(a)). However, as the contact pressure increases, the initiation of C-C
bonding between the a-C and graphene, as well as between graphene and
the IFL, leads to higher shear stress driven by increased resistance to
sliding (approaching shear stress of approximately 8 GPa for the highest
normal pressure). Also, the SiO\textsubscript{2} slab generally exhibits
an increase in shear stress with contact pressure, characterized by a
significant step-like increase towards a shear stress of ca. 6 GPa in
the normal pressures range between 10 GPa and 12.5 GPa. Remarkably, this
transition pressure range is significantly lower for the
SiO\textsubscript{2} than for the a-C case (compare red and black curves
in Fig. 2(b)).

\begin{figure}[htbp]
\centering
\includegraphics[width=0.8\textwidth,height=0.5\textheight,keepaspectratio]{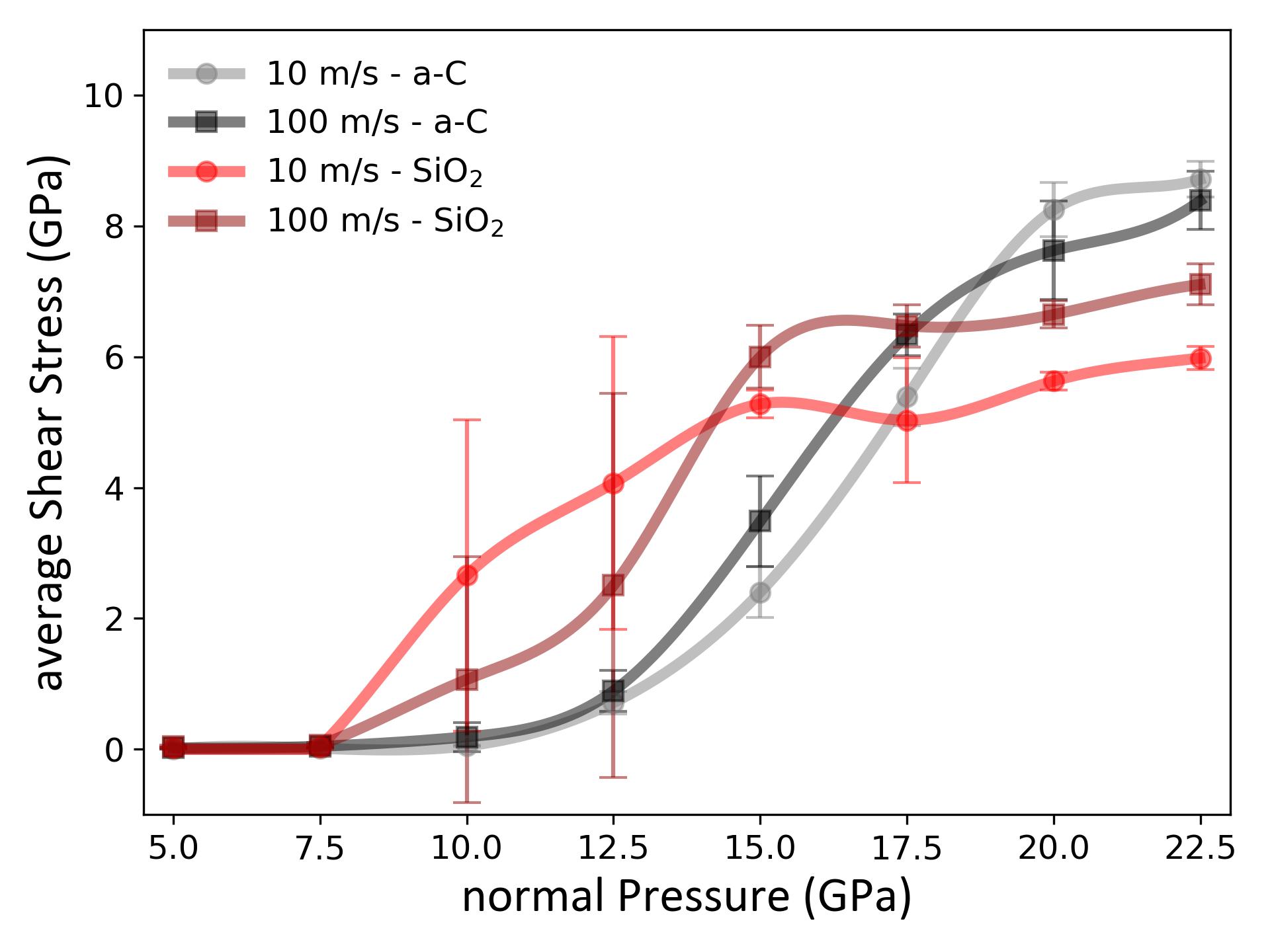}
\caption{Shear stress response of a-C and SiO\textsubscript{2} tips on
graphene/SiC to varying normal pressures and sliding velocities at 300
K.}
\label{fig:3}
\end{figure}

\subsubsection{The Influence of sliding velocity}

The velocity of the FFM tips is orders of magnitude smaller than the
sliding velocities in the DFTB simulations. Therefore, we performed an
additional simulation campaign at a sliding velocity of 10 m/s to assess
the effect of the velocity of the amorphous counter slabs (Fig. 3). For
both, the a-C and SiO\textsubscript{2} counter bodies, the shear stress
values at 10 m/s are close to those at 100 m/s across the range of
normal pressures (5 GPa to 22.5 GPa) and at a temperature of 300 K,
indicating only a marginal influence of sliding velocity. For both
amorphous counter materials, the onset of the transition between regimes
I and II seems to shift to smaller pressures when sliding velocity is
reduced. While sliding velocity has an impact on the transition
pressures, its effect on the shear stress plateau at high normal
pressures is rather weak. See supplementary information for a similar
assessment of the temperature dependence of the shear stress.

\subsubsection{Evolution of total density, oxygen density and velocity
profile}

To further explore interfacial interactions, shear behavior, and
material deformation, we calculated the total density, oxygen density
and velocity profiles under different normal pressures (representing the
various sliding regimes) for different times during the simulations. For
each normal pressure, the data in Fig. 4 is derived from the same sample
as in Fig. 2. The selected trajectory highlights detailed structural and
velocity characteristics, capturing subtle features. To confirm the
representativeness of our choice, we provide an average over all four
independent runs in the supplementary information (Fig. S1). This
averaged data validates the robustness of the trends described in the
main text but obscures finer details that are critical for interpreting
specific interfacial behaviors.

We start with a description of the 10 GPa results. Initially, the total
density (dashed blue curve) shows the expected oscillations of the
crystalline SiC (including the IFL bonded to the SiC) for
\(z < 15\ \mathring{\mathrm{A}}\). The graphene is represented by the
peak at \(z = 17\ \mathring{\mathrm{A}}\), while the a-C gives rise to a
structureless density plateau with 2.2 g/cm\textsuperscript{3} for
\(z \geq 20\ \mathring{\mathrm{A}}\). The oxygen atoms are located at
\(z = 20\ \mathring{\mathrm{A}}\), visible as a pronounced peak in the
initial oxygen density (dashed red curve). The initial velocity profile
(dashed green curve) jumps from 0 m/s to 100 m/s in a narrow region
around \(z = 18\ \mathring{\mathrm{A}}\) indicating sliding of the
H/OH-passivated a-C over the epitaxial graphene. This behavior is
conserved during the whole simulation such that the densities and the
velocity profile at the beginning and the end of the trajectories are
almost identical (compare dashed and solid lines in Fig. 4).

Also, for the 12.5 GPa case, the density profiles did not exhibit any
significant changes during the simulation run. However, in the initial
total density the location of the graphene is lower compared to the 10
GPa case, since bonding between graphene and the IFL starts early in the
simulation. The shear plane shifts downward by approximately
\(1\ \mathring{\mathrm{A}}\) within the first 60 ps (compare dashed and
dot-dashed green curves), reflecting the onset of bonding of the a-C to
the graphene accompanied by slip events of the graphene against the IFL.

As the pressure increases further to 15 GPa and 20 GPa, the depletion of
passivating hydroxyl groups at the a-C/graphene interface results in
cold welding between a-C and the graphene layer. This leads to
substantial chemical mixing of the a-C that drives densification (with
local peaks in the a-C total density reaching almost 3
g/cm\textsuperscript{3}) and migration of oxygen atoms deeper into the
a-C layer (red solid curves). The shear plane moves temporarily into the
bulk of the a-C; see the dash-dotted green curve for the velocity
profile at 60 ps. By 0.2 ns, the chemical mixing has stopped. As
indicated by the solid green velocity profile, the shear plane shifted
toward the interface between the IFL and the SiC substrate.

The total density profiles in Fig. 4 provides a clear signature of
structural transformation within the graphene/IFL zone. At the lowest
pressure (10 GPa), the total density profile in the graphene region
shows two distinct peaks, corresponding to a clear separation between
the graphene and IFL. However, as the pressure increases, these peaks
converge to a bimodal peak. This reflects the increased chemical bonding
between graphene and IFL, finally forming a single layer of diamond (see
Fig. 2(a)).

\begin{figure}[htbp]
\centering
\includegraphics[width=\textwidth,height=0.5\textheight,keepaspectratio]{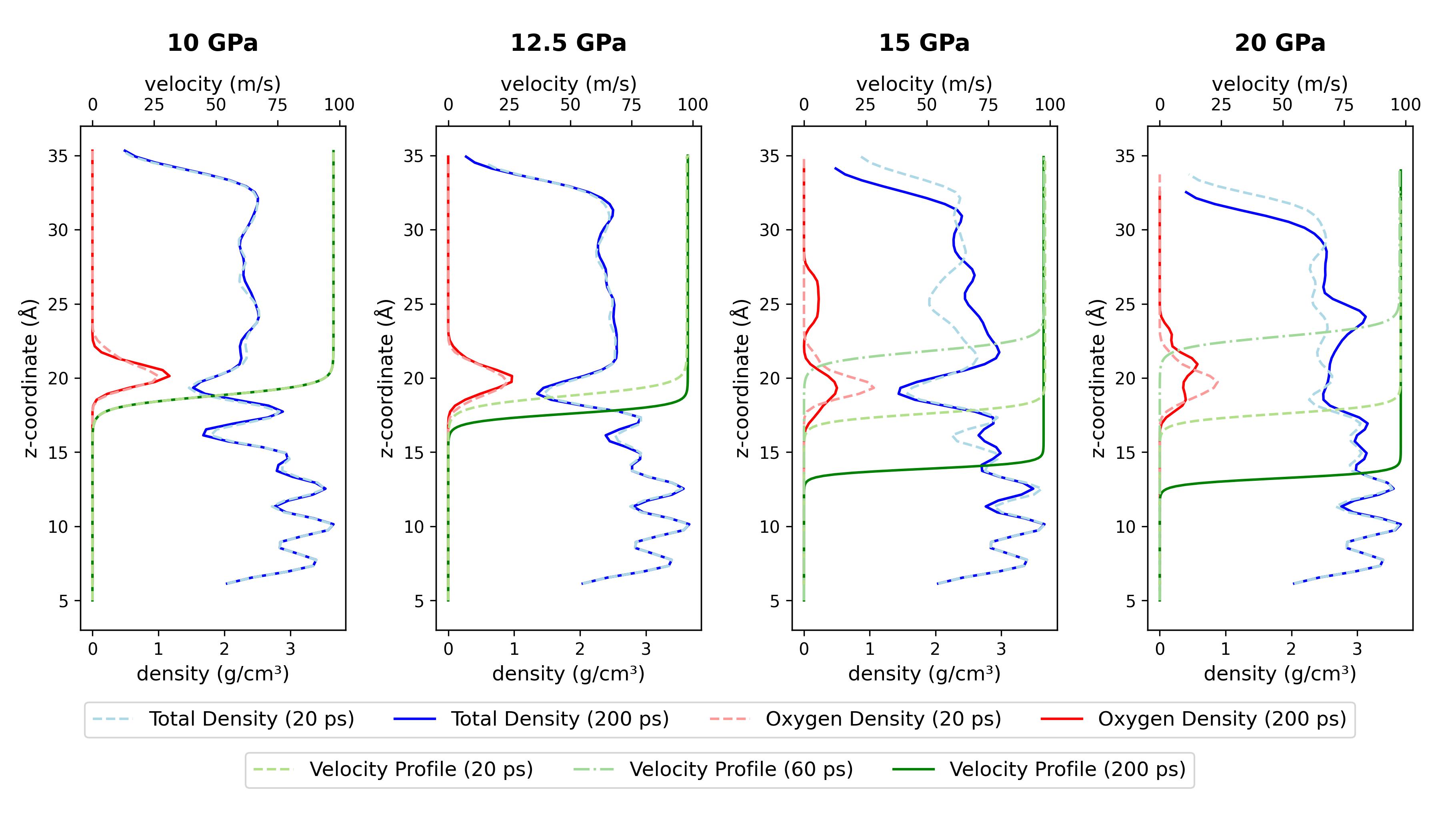}
\caption{Velocity profiles and density distributions of the
a-C/Graphene/SiC system under varying normal pressures (10 GPa, 12.5
GPa, 15 GPa, and 20 GPa) as a function of \(z -\)coordinate. Total
density (blue lines) and oxygen density (red lines) at 20 ps (dashed
lines) and 200 ps (solid lines) illustrate the structural evolution of
the system. Velocity profiles at 20 ps (green dashed line), 60 ps (green
dash-dot line), and 200 ps (solid green line) highlight the location of
the shear plane during this evolution. At higher pressures, intermediate
chemical mixing occurs in the a-C leading to its densification
accompanied by the migration of oxygen atoms into the bulk a-C region.}
\label{fig:4}
\end{figure}

\subsubsection{Bonding Dynamics, System Configuration Evolution and
Shear Stress}

To better understand the evolution of the mechanical strength and
stability of the various interfaces, we analyzed the interfacial bond
formation over time. It is important to note that no new bond formation
was observed at any interface during sliding regime I. Consequently,
this analysis focuses on sliding regimes II and III, where both C-C and
C-Si bond formation was observed. Fig. 5 shows the typical time
evolution of a-C/Graphene/SiC configurations (panel (a) for 12.5 GPa and
panel (b) for 20 GPa normal pressure) for these two regimes. Panels c
and d report the corresponding average numbers of C-C and C-Si bonds
over a 0.2 ns sliding period of the a-C overlayer across the epitaxial
graphene on SiC.

\begin{figure}[htbp]
\centering
\includegraphics[width=1.0\textwidth]{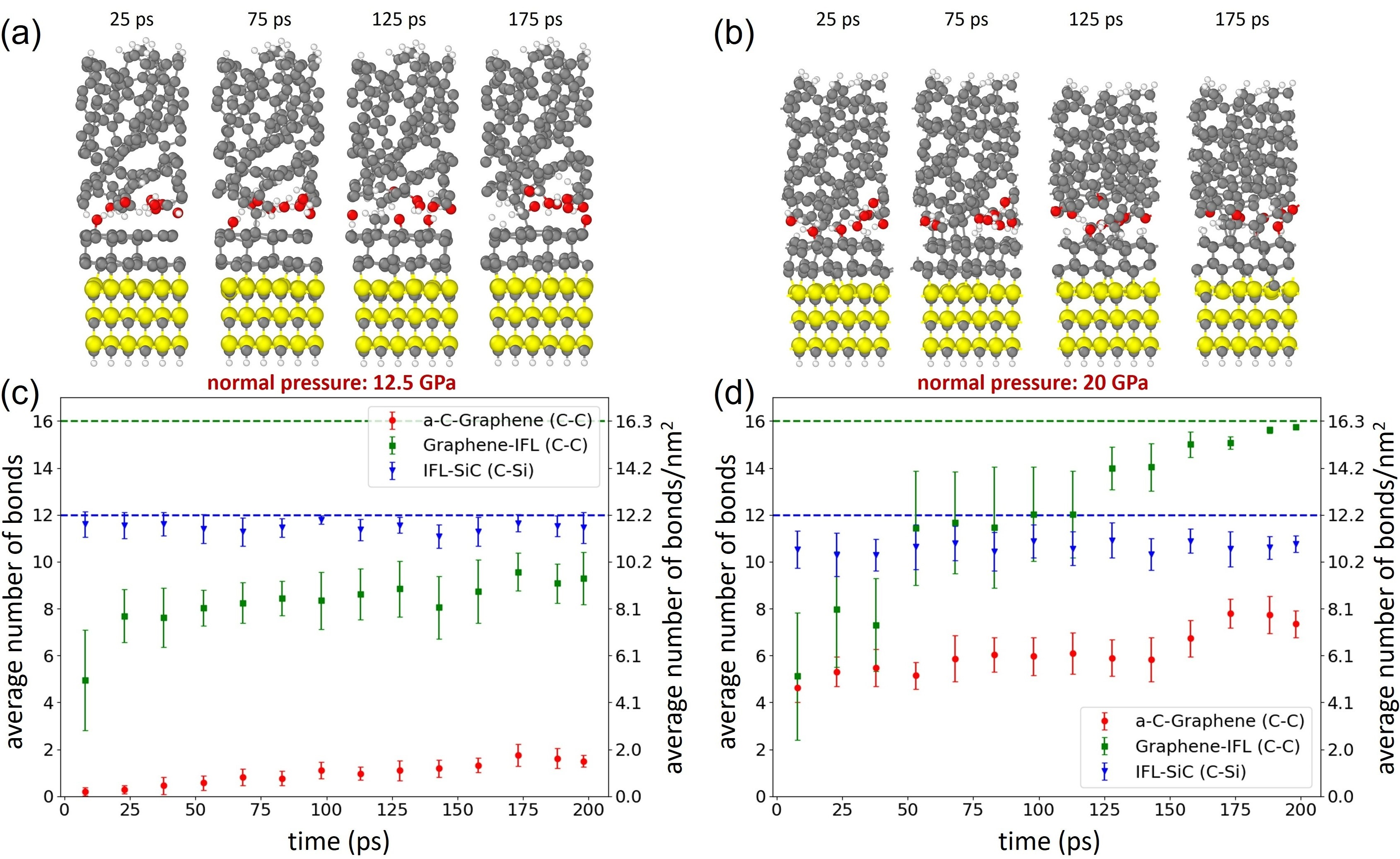}
\caption{Evolution of a-C/Graphene/SiC configurations and corresponding
average number of C-C and C-Si bonds over time for two normal pressures,
12.5 GPa (left column) and 20 GPa (right column), representing regime II
and regime III of sliding, respectively. The top panels a and b display
snapshots of the atomic configurations at different time intervals (25
ps, 75 ps, 125 ps, and 175 ps). The bottom panels c and d illustrate
interactions at three interfaces by reporting the average number of C-C
and C-Si bonds: a-C/graphene (red disks), graphene/IFL (green squares),
and IFL/SiC (blue triangles). Blue and green dashed lines at bond counts
of 12 and 16 represent the maximum possible number of C-Si and C-C bonds
at the IFL/SiC and graphene/IFL interfaces, respectively. The right-hand
\(y\)-axis presents the normalized number of C-C and C-Si bonds per
square nanometer. Each average represents the mean number of bonds over
14 consecutive movie frames, providing a representative value for that
time period. The results were then averaged across the four samples to
obtain a reliable estimate of the mean bond count.}
\label{fig:5}
\end{figure}

In regime II (see Fig. 5(a)), the applied normal pressure is sufficient
to initiate the formation of new C-C bonds between a-C and graphene.
However, the average number of these bonds remains low throughout this
regime, ranging from 0 to approximately 2 bonds per
nm\textsuperscript{2} (see Fig. 5(c)), indicating limited bonding and
interaction between these two layers. Conversely, in regime III (Fig.
5(b)) the number of bonds between a-C and graphene increases to 8 bonds
per nm\textsuperscript{2}. Apparently, the densification of the a-C
brings more carbon atoms into contact with the graphene. This results in
greater frictional resistance, leading to increase in shear stress (Fig.
2(b)). Even with just 8 C-C bonds/nm\textsuperscript{2}, the a-C layer
can effectively cold weld with the graphene/IFL single layer diamond,
highlighting the critical role of these bonds in enhancing the overall
mechanical behavior under elevated pressure conditions.

At the graphene/IFL interface, the C-C bond count shows an even more
pronounced dependence on the normal pressure. In regime II (see Fig.
5(c)), the bond count starts around 5 bonds per nm\textsuperscript{2}
and gradually increases to about 9 bonds/nm\textsuperscript{2} over
time, which is significantly below the possible maximum of 16.3
bonds/nm\textsuperscript{2} (indicated by the green dashed line). This
gradual increase indicates the activation of additional bonding sites
between the graphene and IFL during sliding. In regime III (Fig. 5(d)),
the bond count between graphene and the IFL rises further, approaching
the maximum possible count of 16 bonds/nm\textsuperscript{2}. This
complete bond formation at the graphene/IFL interface suggests a strong
structural transformation resulting in the formation of a single layer
of diamond.

Finally, we consider the number of bonds at the IFL/SiC interface.
During regime II (Fig. 5(c)), the bond count remains consistently high,
around the theoretical maximum of 12 bonds/nm\textsuperscript{2}.
Occasional fluctuations can be attributed to slip events between the IFL
and the SiC substrate. In regime III (Fig. 5(d)), a different scenario
unfolds. The increased bonding between graphene and the IFL promotes the
formation of the single layer diamond, whose cold welding with the a-C
causes the IFL to slip at a sliding velocity of 100 m/s over the SiC
surface. This results in the successive breaking and reformation of C-Si
bonds at the IFL/SiC interface, as reflected in the bond count
fluctuation between 10 and 12 bonds/nm\textsuperscript{2} (Fig. 5(d)).

\begin{figure}[htbp]
\centering
\includegraphics[width=0.8\textwidth,height=0.5\textheight,keepaspectratio]{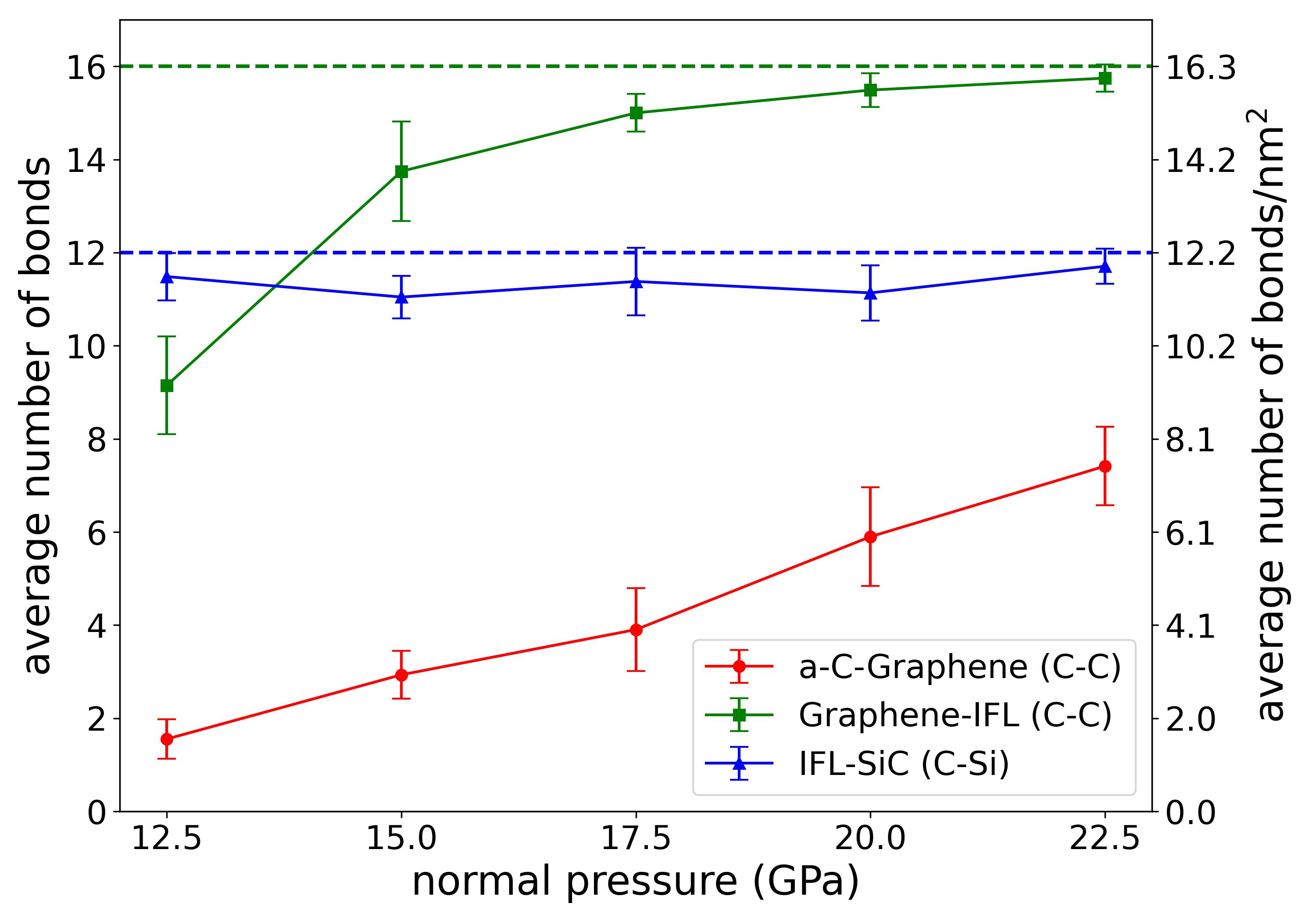}
\caption{Average number of bonds as a function of normal pressure for
three different interfaces: a-C/graphene (C-C), graphene/IFL (C-C), and
IFL/SiC (C-Si), based on data from the last 50 ps of the simulations.
The dashed lines indicate the possible maximum number of bonds for the
graphene/IFL (16 bonds, green) and IFL/SiC (12 bonds, blue) interfaces.
The secondary y-axis on the right shows the normalized bond count per
square nanometer.}
\label{fig:6}
\end{figure}

To understand how interfacial interactions influence changes in shear
stress and mechanical stability, we studied the effect of varying normal
pressures on the final bond density at different interfaces. Fig. 6
displays the average number of C-C bonds at the a-C/graphene and
graphene/IFL interface as well as the average number of C-Si bonds at
the IFL/SiC interface as a function of normal pressure. At the
a-C/graphene interface, the number of C-C bonds gradually increases with
the normal pressure, from approximately 2 bonds/nm\textsuperscript{2} at
12.5 GPa to approximately 7 bonds/nm\textsuperscript{2} at 22.5 GPa.
Note, that this increase in bond formation correlates directly with the
rise in shear stress observed at this interface (Fig. 2(b)) - as more
C-C bonds form between the a-C and graphene layers, the interfacial
adhesion is enhanced, leading to increased frictional resistance and,
consequently, higher shear stress.

For the graphene/IFL interface, the number of C-C bonds increases
rapidly as the normal pressure rises, approaching the possible maximum
of 16 bonds/nm\textsuperscript{2}. This indicates that at higher
pressures, this interface achieves near-complete bond saturation, which
also contributes to the overall increase in shear stress and mechanical
stability. In contrast, the IFL/SiC interface exhibits remarkable
stability across all normal pressures, with the number of C-Si bonds
fluctuating slightly below the possible maximum of 12
bonds/nm\textsuperscript{2}. This consistent bond count, despite the
dynamic adjustments observed earlier, suggests that the interface is
able to maintain its structural integrity even under increased
pressures. This stability implies that the IFL/SiC interface plays a
supporting role in maintaining mechanical stability, without
significantly impacting the variations in shear stress observed at other
interfaces. Consequently, while the a-C/graphene and graphene/IFL
interfaces show a direct correlation between bond density and shear
stress, the IFL/SiC interface provides a stable foundation that adapts
to the shifting shear plane, accommodating bond breaking and reformation
activities at this interface without compromising its overall bond
density.

\subsubsection{Structure of the epitaxial graphene after sliding}

On the one hand, the simulations showed extensive bonding between the
IFL and graphene as well as between the graphene and the a-C under
regime II and III conditions resulting in complete cold welding of the
entire tribo system (consisting now of SiC/single layer diamond/a-C cold
welded layer system). On the other hand, the experimental post-sliding
analysis revealed that the distance between the graphene and IFL layers
returned to the typical spacing of two graphene layers after sliding
(i.e. no trace of the single layer diamond or adatoms on the graphene).
This opens the question if the simulations also predict a recovery of
the graphene to its original state after the sliding simulations. To
elucidate the structure of the epitaxial graphene after tip retraction,
we conducted additional simulations where the a-C layer was lifted
upwards after the initial 0.2 ns sliding period. During these
simulations, the a-C was slid in the \(x\)-direction at a \(x\)-velocity
of 100 m/s while being gradually retracted in the normal direction at a
\(z\)-velocity of 10 m/s. Fig. 7 illustrates this process by a series of
snapshots from the 0.2 ns lift-off simulation for normal pressures of 15
GPa and 17.5 GPa.

At 15 GPa, the snapshots in the top row of Fig. 7 reveal the gradual
detachment of the a-C from the graphene interface during the lift-off
process. Initially, one oxygen and one hydrogen atom are bonded to the
single layer diamond and few C-C bonds bridge the gap between the single
layer diamond and the a-C (Fig. 7(a)). As the retraction of the a-C
progresses, these C-C bonds and a series of C-C bonds in the single
layer diamond are broken (Fig. 7(b)). Simultaneously, a free CO molecule
forms, while a hydroxyl group, a CH group, and a hydrogen atom remain
attached to the graphene surface. In principle, the OH group and one of
the hydrogen atoms could combine to form a water molecule, while the
remaining carbon adatom could react with another oxygen atom on the
surface or the tip to produce a second CO molecule. Given that the
experimental tip is significantly larger than the simulated one and
operates at a much lower sliding speed, the reaction of the hydrogen and
carbon atoms on the graphene surface with the tip, followed by their
removal, is highly probable. This process would ultimately lead to the
opening of the remaining single-layer diamond bonds, allowing the
graphene to recover its structure and return to its initial distance
from the IFL.

\begin{figure}[htbp]
\centering
\includegraphics[width=0.8\textwidth]{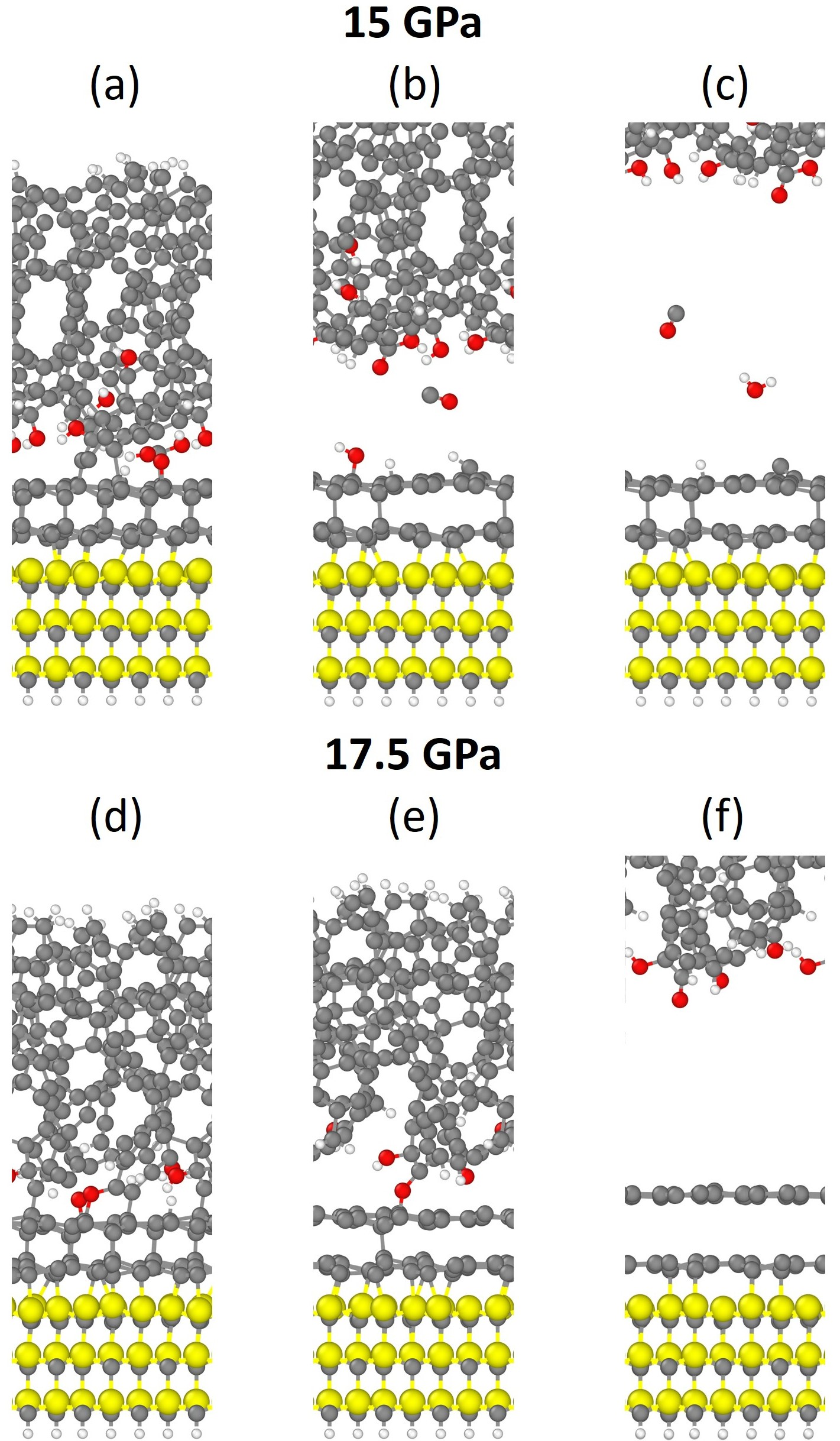}
\caption{Snapshots from the lift-off simulations of the a-C tip following
0.2 ns of simulations at normal pressures of 15 GPa (top row) and 17.5
GPa (bottom row). The frames arranged sequentially from left to right,
illustrate key stages of the structural evolution during the lift-off
phase. The images highlight the restoration of interlayer distances and
changes in interfacial configurations, providing insights into the
system\textquotesingle s behavior during retraction.}
\label{fig:7}
\end{figure}

To further analyze this behavior, we performed DFT calculations to
evaluate the energetics of the desorption of H and OH from the graphene
surface and the subsequent formation of a water molecule in the tribo
gap (Fig. 7(c)). The energy difference between the system configurations
in Fig. 7(b) and Fig. 7(c) is negative ($E_{\text{panel c}} - E_{\text{panel b}} = -0.1378 < 0$), indicating that the
formation of a water molecule is energetically favorable compared to
having H and OH adsorbed on the surface. This result highlights the
thermodynamic preference for water formation and its detachment from the
graphene interface. With the same reasoning, one can argue that the
remaining H and C adatoms will leave the graphene, resulting in an
incomplete single layer diamond free of adatoms. Finally, heating this
system to 500 K demonstrates that the graphene fully detaches from the
IFL.

At 17.5 GPa, the situation is even clearer, as the lift-off proceeds
through the breaking of various C-C and C-O bonds (see the evolution
from Fig. 7(d) to Fig. 7(e)) between the graphene and the a-C layer.
This process is accompanied by the dissociation of nearly all interplane
bonds in the single-layer diamond structure. After the final ether
bridge breaks (between Fig. 7(e) and Fig. 7(f)), the ideal epitaxial
graphene spontaneously reforms. The behavior observed in these lift-off
simulations suggests that the graphene interface is capable of returning
to its pre-sliding configuration, demonstrating the resilience of the
system under moderate pressure and thermal conditions. This finding is
consistent with the experimental results.

\subsubsection{Effect of the IFL mobility}

As shown in Fig. 2(a), during sliding regime III, the single-layer
diamond (formed, for example, at 20 GPa) is dragged along with the a-C.
However, this scenario is unlikely in experiments since the graphene
flakes are micron-sized and pushing them over the SiC surface would
require unrealistically large forces. In this context, the periodic
boundary conditions in the simulations translate into an infinite array
of tips capable of shifting entire graphene flakes---an artifact of the
simulation setup.

\begin{figure}[htbp]
\centering
\includegraphics[width=\textwidth,height=0.5\textheight,keepaspectratio]{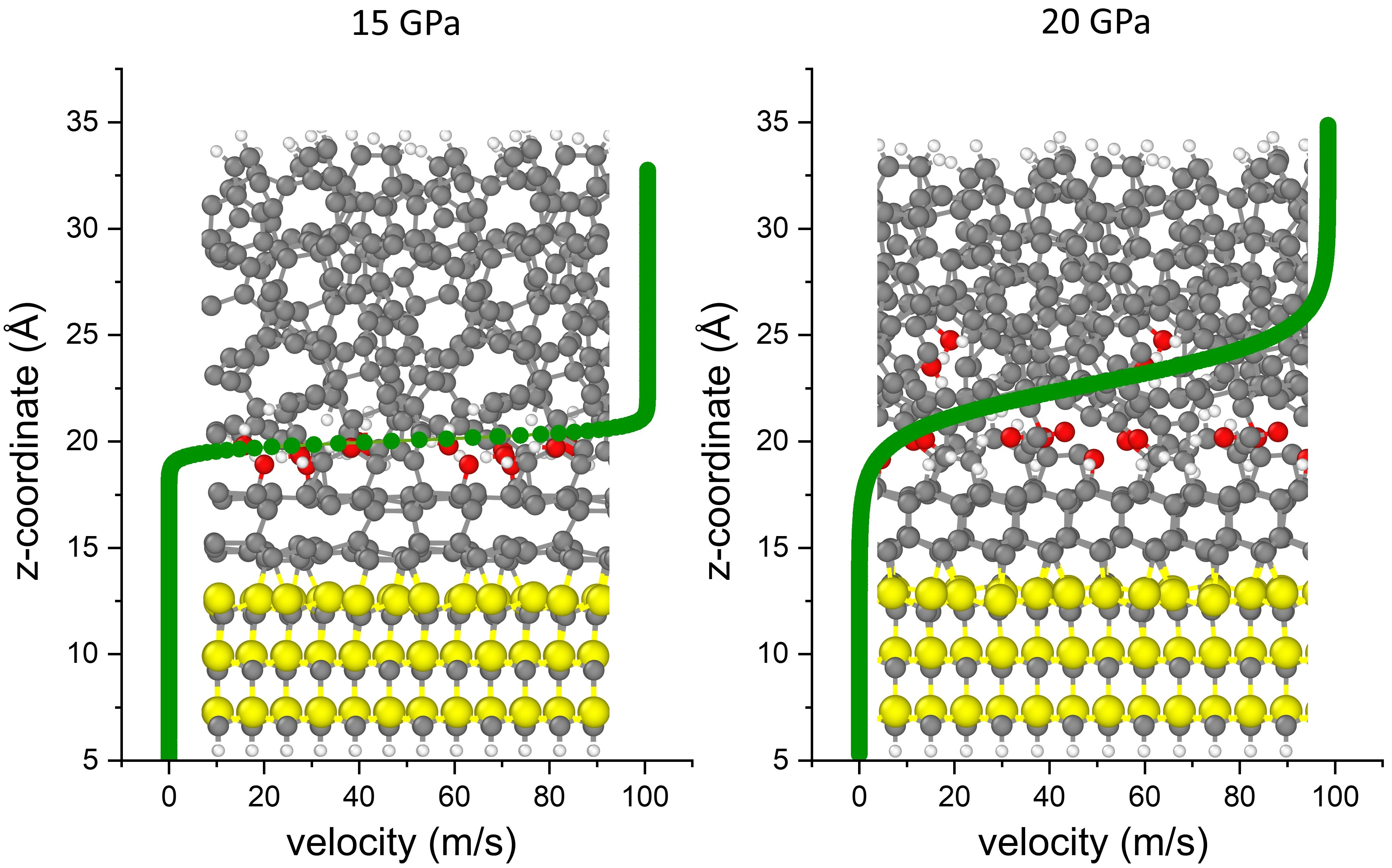}
\caption{Velocity profile illustrating the shear plane shift into the a-C
region after freezing the x-position of the IFL atoms during the
modified 20 GPa simulation.}
\label{fig:8}
\end{figure}

To address this issue, we took the samples from normal pressures 15 GPa
and 20 GPa simulations after 0.2 ns and made a minor modification to the
protocol. Specifically, we froze the x-position of the IFL atoms to
ensure that the single-layer diamond remained anchored to the SiC. This
adjustment mitigated the artifact caused by periodicity in the
simulation setup. We then ran the sliding simulation for an additional
0.2 ns.

The resulting velocity profiles, shown in Fig. 8, illustrate a shift in
the shear plane location into the a-C region for both normal pressures.
The fixed IFL layer, combined with multiple (a-C)C-C(graphene) bonds
that anchor the a-C atoms near the graphene layer to the interface,
causes the shear plane to relocate upward into the a-C layer. At 15 GPa,
the shear plane is located in the a-C, close to the graphene layer,
while at 20 GPa, the shear plane is located deeper within the a-C layer.
The average steady shear stress increased significantly with the
modified setup, rising from \(\tau = 3.5\ GPa\) to \(\tau = 7.9\ GPa\)
for normal pressure 15 GPa and from \(\tau = 7.6\ GPa\) to
\(\tau = 12.3\ GPa\) for normal pressure 20 GPa. This result aligns with
prior studies on the plasticity of amorphous carbon ~{[}13{]}, showing
that the interface sets a lower bound on the flow stress, with shear
stress increasing linearly above 8 GPa, indicating enhanced resistance
to shear under compression.

\section{CONCLUSIONS}

This study investigates the sliding interaction between an a-C covered
diamond tip and an epitaxial graphene layer on SiC(0001), comparing it
with the behavior of a SiO\textsubscript{2} covered Si tip. Experimental
measurements and molecular dynamics simulations using the DFTB method
were conducted to explore the tribological properties of these materials
under various normal pressures and temperatures.

The experimental results show that for the Si tip, an initial linear
low-friction regime is observed up to a normal force of 70 nN, followed
by a step-like increase in friction. High-friction values generally
increase with pressure but exhibit significant scatter, occasionally
returning to the low-friction regime. For the diamond tip, a
low-friction regime with fluctuating friction force is observed for
normal forces up to 400 nN, followed by a linear-like increase in
friction force. These experimental observations align with the
simulation results.

MD simulations identified distinct sliding regimes, each characterized
by unique bond formation dynamics and shear plane locations. At low
pressures, no chemical bonds formed, and the shear plane was located
between the tip and graphene (Regime I: \textless12.5 GPa for a-C,
\textless10 GPa for SiO\textsubscript{2}). As the pressure increased,
a-C--Graphene and SiO\textsubscript{2}--Graphene bond formation
commenced, while the shear plane remained between the tips and graphene
(Regime II: 12.5 GPa to \textless15 GPa for a-C, 10 GPa to \textless15
GPa for SiO\textsubscript{2}). At higher pressures, the behavior of the
two materials diverged. For a-C, tip-graphene bond formation was
accelerated, with the shear plane primarily located between the IFL and
SiC. In contrast, for SiO\textsubscript{2}, the shear plane shifted
upward into the silicon oxide (Regime III: 15 GPa to $\leq$22.5 GPa for both
a-C and SiO\textsubscript{2}).

This study highlights the distinct behaviors of a-C and
SiO\textsubscript{2} in frictional interactions with graphene, driven by
their material properties and bonding mechanisms. The a-C tip exhibits
robust C-C bonding and rigidity, while SiO\textsubscript{2} is
relatively softer and allows for greater deformation. This results in
different shear plane locations and frictional responses under similar
pressure conditions.

In sliding regimes II and III, the time evolution of bond formation and
system configuration was closely linked to changes in mechanical
behavior and shear stress. At lower pressures in regime II, limited bond
formation between a-C and graphene resulted in lower shear stress. At
higher pressures in regime III, the increased number of C-C bonds
contributed to higher frictional resistance and enhanced mechanical
stability. The IFL/SiC interface remained resilient under both
conditions. In regime II, minor fluctuations occurred due to slight
bouncing. In regime III, despite dynamic C-Si bond changes, the
interface maintained a high bond count, showcasing its adaptability and
stability. As the normal pressure increases from 12.5 to 22.5 GPa, C-C
bonds at the a-C/graphene interface steadily rise, enhancing adhesion
and shear stress, while at the graphene/IFL interface, bond formation
rapidly reaches near-maximum levels, indicating strong bonding and
increased frictional resistance. In contrast, the stable bond count at
the IFL/SiC interface suggests minimal impact on shear stress
variations.

Temperature and sliding velocity studies further demonstrate that a-C
exhibits a significant increase in shear stress with temperature and
pressure due to enhanced atomic activity and bonding interactions. In
contrast, SiO\textsubscript{2} shows minimal changes in shear stress
across varying conditions, which is attributed to its softer material
properties and stable interfacial interactions.

These findings underscore the differing behaviors of a-C and
SiO\textsubscript{2} in frictional interactions with graphene and
provide valuable insights for the design of low-friction materials and
coatings.

\section{ACKNOWLEDGMENTS}

We acknowledge financial support by Deutsche Forschungsgemeinschaft
within the Research Unit 5099 (M.M. and M.Z.) and within the Priority
Program SPP 2244 ``2DMP'' (B.S. and R.B.). Sample preparation by Thomas
Seyller (University of Technology Chemnitz) is gratefully acknowledged.
Computing time was granted by the John von Neumann Institute for
Computing (NIC) and provided on the supercomputer JUWELS at Jülich
Supercomputing Centre (JSC) within the Project Harsh.

\section{REFERENCES}

{[}1{]} D. Berman, A. Erdemir, and A. V. Sumant, Graphene: A new
emerging lubricant, Mater. Today \textbf{17}, 31 (2014).

{[}2{]} S. Kawai et al., Superlubricity of graphene nanoribbons on gold
surfaces, Science (80-. ). \textbf{351}, 957 (2016).

{[}3{]} C. Lee, Q. Li, W. Kalb, X. Z. Liu, H. Berger, R. W. Carpick, and
J. Hone, Frictional characteristics of atomically thin sheets, Science
(80-. ). \textbf{328}, 76 (2010).

{[}4{]} X. Ge, Z. Chai, Q. Shi, Y. Liu, and W. Wang, Graphene
superlubricity: A review, Friction \textbf{11}, 1953 (2023).

{[}5{]} T. Filleter, J. L. McChesney, A. Bostwick, E. Rotenberg, K. V.
Emtsev, T. Seyller, K. Horn, and R. Bennewitz, Friction and dissipation
in epitaxial graphene films, Phys. Rev. Lett. \textbf{102}, 1 (2009).

{[}6{]} B. Szczefanowicz, T. Kuwahara, T. Filleter, A. Klemenz, L.
Mayrhofer, R. Bennewitz, and M. Moseler, Formation of intermittent
covalent bonds at high contact pressure limits superlow friction on
epitaxial graphene, Phys. Rev. Res. \textbf{5}, 1 (2023).

{[}7{]} K. V. Emtsev et al., Towards wafer-size graphene layers by
atmospheric pressure graphitization of silicon carbide, Nat. Mater.
\textbf{8}, 203 (2009).

{[}8{]} M. Elstner, D. Porezag, G. Jungnickel, J. Elsner, M. Haugk, S.
Suhai, and G. Seifert, Self-consistent-charge density-functional
tight-binding method for simulations of complex materials properties,
\textbf{58}, 7260 (1998).

{[}9{]} \emph{Atomistica Software Suite},
https://github.com/Atomistica/atomistica.

{[}10{]} L. Pastewka, S. Moser, and M. Moseler, Atomistic Insights into
the Running-in , Lubrication , and Failure of Hydrogenated Diamond-Like
Carbon Coatings, \textbf{39}, 49 (2010).

{[}11{]} D. Frenkel and B. Smit, \emph{Understanding Molecular
Simulation} (Academic Press, San Diego, 2002).

{[}12{]} F. P. Bundy, W. A. Bassett, M. S. Weathers, R. J. Hemley, H. K.
Mao, and A. F. Goncharov, The pressure-temperature phase and
transformation diagram for carbon; updated through 1994, Carbon N. Y.
\textbf{34}, 141 (1996).

{[}13{]} R. Jana, J. von Lautz, S. M. Khosrownejad, W. B. Andrews, M.
Moseler, and L. Pastewka, Constitutive relations for plasticity of
amorphous carbon, JPhys Mater. \textbf{3}, 035005 (2020).

\section{SUPPLEMENTARY INFORMATION}

\includegraphics[width=\textwidth,height=0.5\textheight,keepaspectratio]{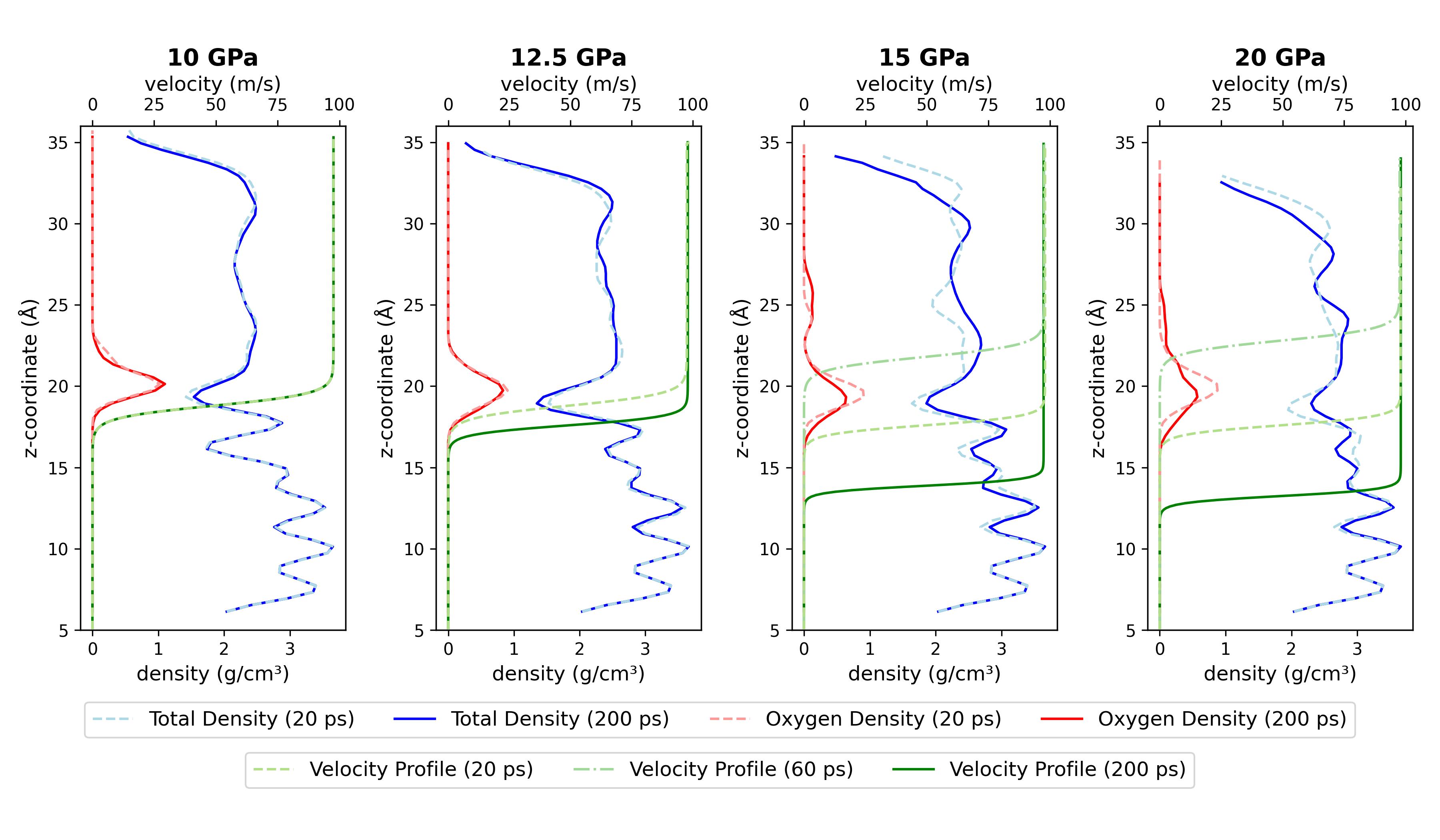}\vspace{1em}

\textbf{Figure S1.} The averaged velocity profiles and density
distributions of the a-C/Graphene/SiC system under varying normal
pressures (10 GPa, 12.5 GPa, 15 GPa, and 20 GPa) as a function of the
z-coordinate. Total density (solid blue line) and oxygen density (solid
red line) illustrate the structural evolution of the system. Velocity
profiles at 20 ps (green dashed line), 60 ps (green dash-dot line), and
200 ps (solid green line) indicate the location of the shear plane.\vspace{1em}

\emph{\textbf{Temperature dependence}}\vspace{1em}

The effect of temperature on shear stress is investigated for a-C and
compared to that of SiO\textsubscript{2} slabs interacting with graphene
surfaces under increasing normal pressures at temperatures of 300 K, 500
K, and 1000 K, as shown in Fig S2. Our results indicate that
a-C\textquotesingle s shear stress significantly increases with both
temperature and pressure. At lower contact pressures, a-C slides
smoothly over graphene with marginal shear stress due to the absence of
chemical bonding. However, as normal pressure increases, bonding between
graphene layers and a-C leads to higher shear stress. Trajectory
analysis at low pressures reveals that with increasing temperature,
bonding between a-C and graphene commences at 10 GPa at both 500 K and
1000 K, which is lower than the initiation point of 12.5 GPa observed at
300 K. Elevated temperatures cause thermal expansion and increased
atomic activity within the a-C, enhancing interactions with the graphene
surface and further raising shear stress, as corroborated by the
increased number of C-C bonds at a-C/graphene and graphene/IFL
interfaces observed in Fig S3. This effect is particularly evident at
normal pressures up to 17.5 GPa where the system integrity is not
compromised by the increasing temperature. At higher temperatures,
graphene\textquotesingle s structural integrity is weakened due to
increased thermal activity, making it more susceptible to transformation
under pressure.~Trajectory analysis reveals a distinctive phenomenon
that emerges with rising temperature. Specifically, at 500 K and 1000 K,
graphene layers begin to amorphize under normal pressures of 22.5 GPa
and 20 GPa, respectively.\vspace{1em}

\begin{center}
  \includegraphics[width=0.75\textwidth, height=0.5\textheight, keepaspectratio]{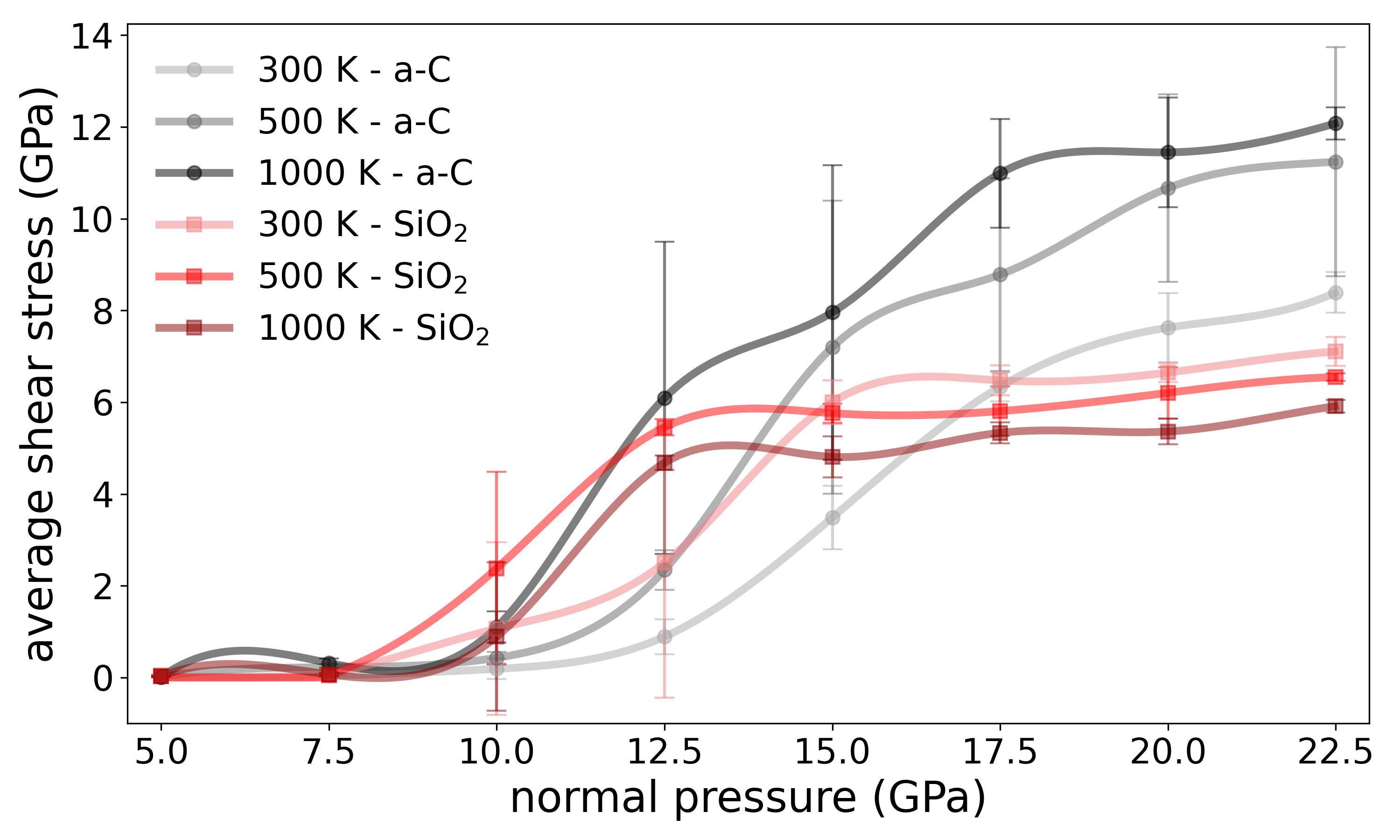}
\end{center}

\textbf{Figure S2.} Shear stress as a function of normal pressure for
a-C and SiO\textsubscript{2} tips sliding over a graphene/SiC system at
different temperatures (300 K, 500 K, and 1000 K) for sliding velocity
100 m/s. a-C shows increasing shear stress with temperature, while
SiO\textsubscript{2} exhibits negligible change with temperature.\vspace{1em}

\includegraphics[width=\textwidth,height=0.5\textheight,keepaspectratio]{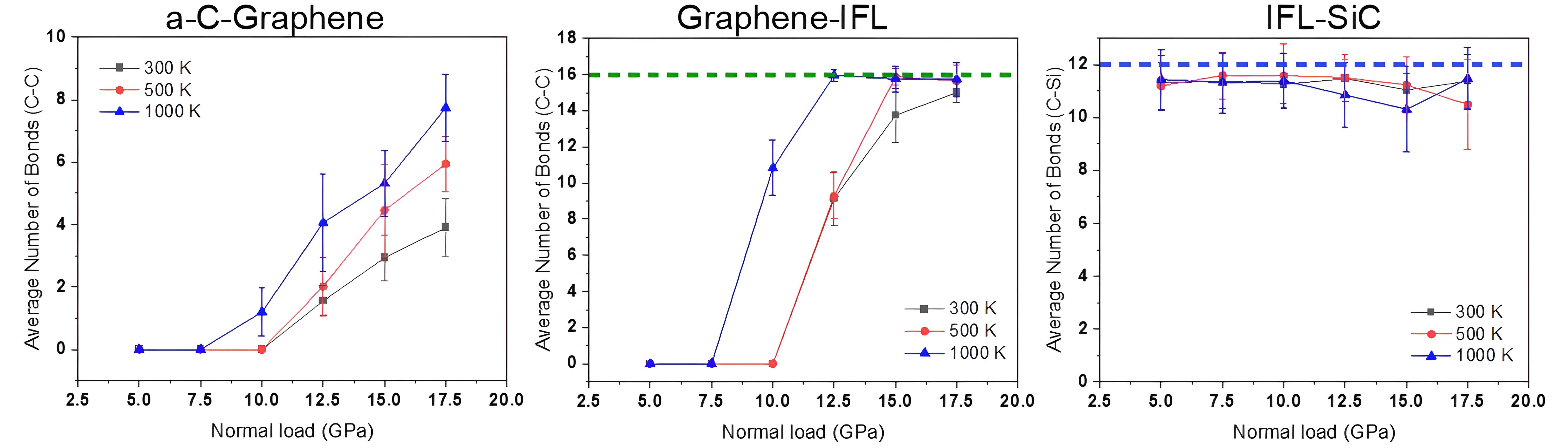}

\textbf{Figure S3.} Average number of bonds formed under varying normal
pressures (5-17.5 GPa) at different temperatures (300 K, 500 K, 1000 K)
for three different interfaces: a-C/Graphene (C-C), Graphene/IFL (C-C),
and IFL/SiC (C-Si).\vspace{1em}

In contrast, SiO\textsubscript{2} exhibits different behavior. Due to
the nature of SiO\textsubscript{2}, it experiences trivial changes in
shear stress across various temperatures and normal pressures. The
relatively softer nature of SiO\textsubscript{2}, compared to a-C,
allows for greater deformation and significant Si-C bond formation with
the graphene layer. As previously mentioned, the differing behaviors of
a-C and SiO\textsubscript{2} tips under varying conditions are
attributed to the types of bonds formed and their material properties.

\end{document}